\documentclass[fdp,fleqn]{w-art}
\usepackage{times,amssymb}
\usepackage[retainorgcmds]{IEEEtrantools}
\usepackage{w-thm}
\usepackage[]{graphicx}
\begin{document}
\DOIsuffix{theDOIsuffix}
\Volume{55}
\Issue{1}
\Month{01}
\Year{2007}
\pagespan{3}{}
\Receiveddate{} \Reviseddate{} \Accepteddate{} \Dateposted{} 



\title[]{Probing CP Violation with Kinematic Reconstruction at the LHC}


\author[K. Rolbiecki]{Krzysztof Rolbiecki\inst{1}%
  \footnote{Corresponding author\quad E-mail:~\textsf{krzysztof.rolbiecki@durham.ac.uk}
}}
\author[G. Moortgat-Pick]{Gudrid Moortgat-Pick\inst{1}}
\address[\inst{1}]{IPPP, University of Durham, Durham DH1 3LE, United Kingdom}
\author[J. Tattersall]{Jamie Tattersall\inst{1}}
\author[P. Wienemann]{Peter Wienemann\inst{2}}
\address[\inst{2}]{Department of Physics, University of Bonn, Nussallee 12, D-53115 Bonn, Germany}
\begin{abstract}
We discuss the potential of observing effects of CP-violation phases
in squark decay chains at the LHC~\cite{MoortgatPick:2009jy}. As the
CP-odd observable, we use the asymmetry composed by triple products
of final state momenta. There are good prospects of observing these
effects using the method of kinematic reconstruction for the final
and intermediate state particles. We also discuss the main
experimental factors and the expected sensitivity.
\end{abstract}
\maketitle                   





\section{Introduction}

The search for Supersymmetry (SUSY) is one of the main goals of
present and future colliders since it is one of the best motivated
extensions of the Standard Model (SM). An important feature of SUSY
models is the possibility of introducing many new sources of CP
violation as required to explain baryon asymmetry in the universe. A
careful analysis of how to observe new CP-violating effects in
future experiments will be required and in the following we discuss
the example in the Minimal Supersymmetric Standard Model.

CP-odd observables are the unambiguous way of discovering hints of
complex parameters in the underlying theory. One of the examples of
such an observable is via exploiting triple product correlations of
momenta and/or spins of the final state particles, see
\cite{Kittel:2009fg} for a recent review. They follow from totally
antisymmetric expressions $i\epsilon_{\mu\nu\rho\sigma} a^{\mu}
b^{\nu} c^{\rho} d^{\sigma}$, where $a$, $b$, $c$ and $d$ are
4-momenta or spins of the particles involved. Such a covariant
product can be expanded in terms of the momenta as follows
\begin{eqnarray}
  \epsilon_{\mu\nu\rho\sigma}p_a^\mu p_b^\nu p_c^\rho p_d^\sigma = &
  E_a\;\overrightarrow{p_b}\cdot(\overrightarrow{p_c}\times\overrightarrow{p_d})+E_c\;
  \overrightarrow{p_d}\cdot(\overrightarrow{p_a}\times\overrightarrow{p_b}) \nonumber \label{eq:EpsExpanLab} \\
  & -E_b\;\overrightarrow{p_c}\cdot(\overrightarrow{p_d}\times\overrightarrow{p_a})
  -E_d\;\overrightarrow{p_a}\cdot(\overrightarrow{p_b}\times\overrightarrow{p_c})\;.
\end{eqnarray}
As we evaluate only one triple product, we miss the contributions of
the other combinations and the asymmetry is diluted. However, if we
are in the rest frame of the decaying particle, the momentum
components of the four vector vanish and we are now only left with
the single triple product that we are interested in:
\begin{equation}
  \epsilon_{\mu\nu\rho\sigma}p_a^\mu p_b^\nu p_c^\rho p_d^\sigma \longrightarrow  m_a\;
  \overrightarrow{p_b}\cdot(\overrightarrow{p_c}\times\overrightarrow{p_d})\;.
 \label{eq:EpsExpanRest}
\end{equation}

\begin{figure}
\begin{minipage}{200pt}
\includegraphics[width=140pt,angle=270]{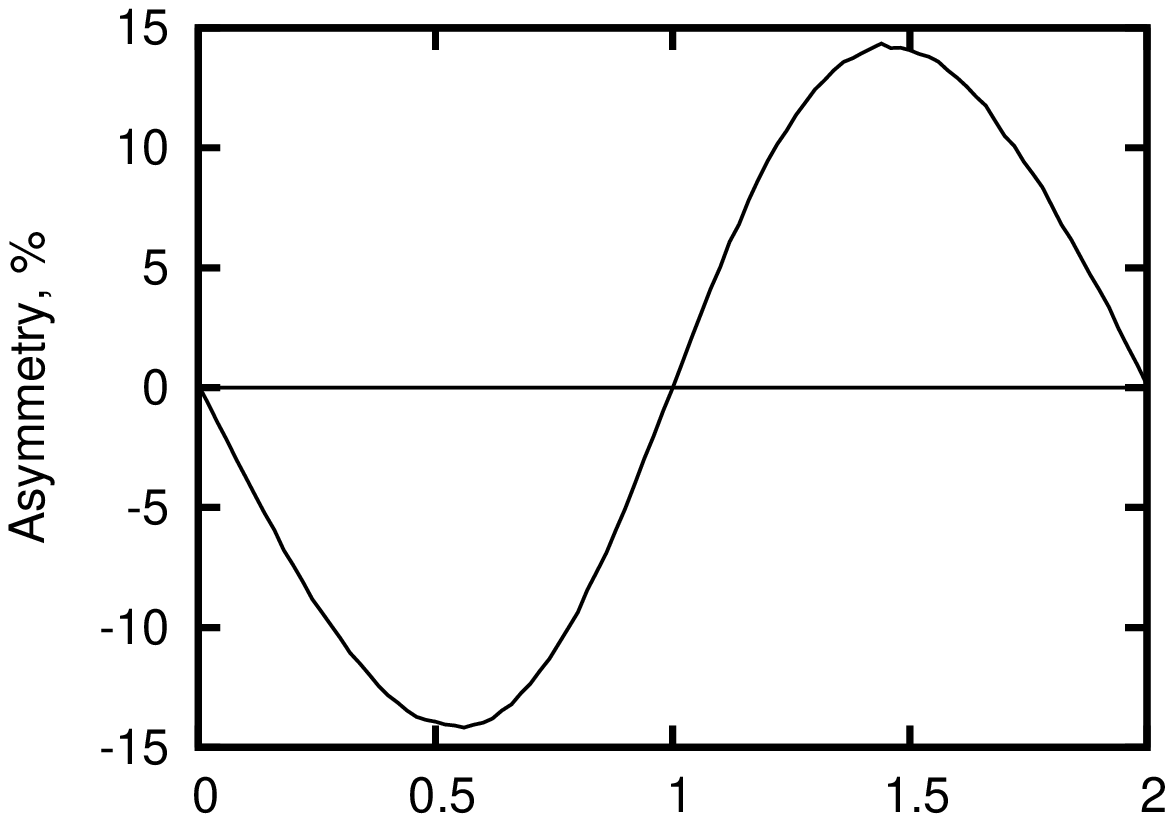}\caption{The
asymmetry $\mathcal{A}_T$, Eq.~(\ref{eq:asy}), in the rest frame of
$\tilde{\chi}^0_2$ as a function of $\phi_1$.\label{fig:asymmetry}}
\end{minipage}\hfill
\begin{minipage}{200pt}
\includegraphics[width=140pt,angle=270]{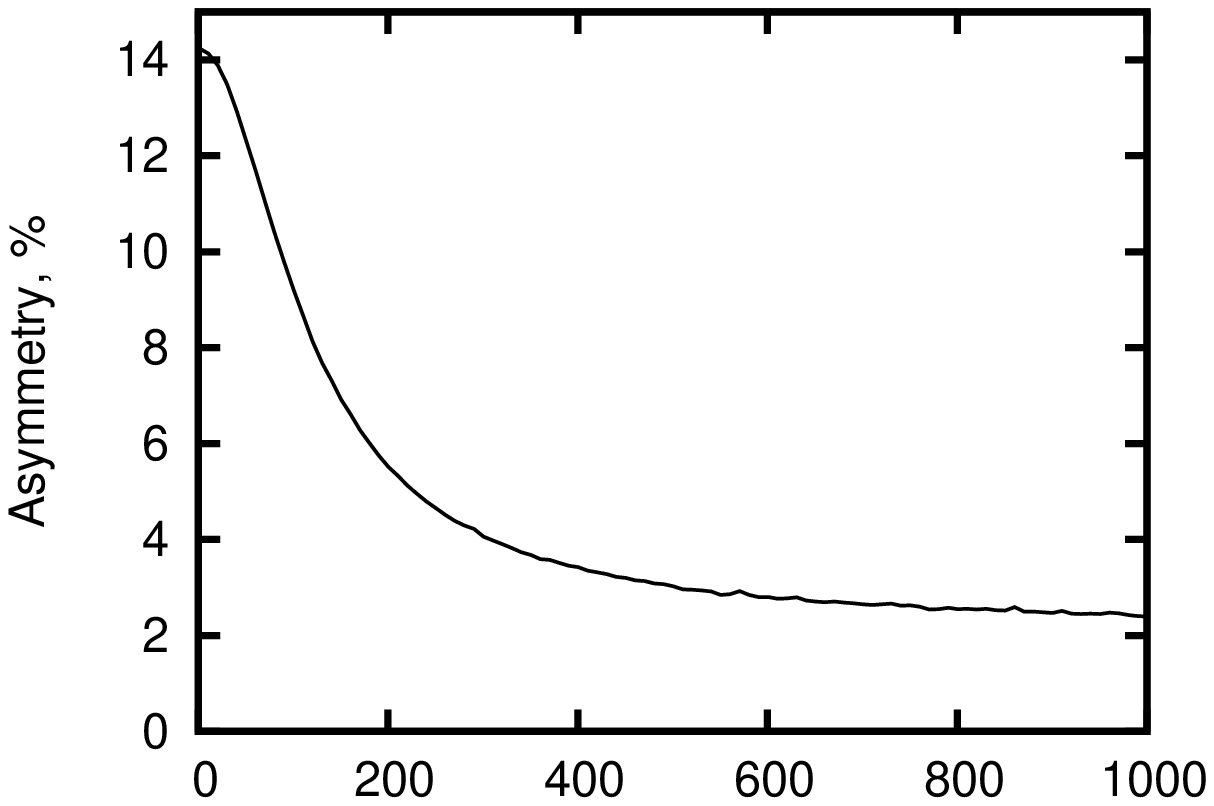} \caption{The
asymmetry $\mathcal{A}_T$, Eq.~(\ref{eq:asy}), in the laboratory
frame as a function of the squark momentum,
$|\vec{p}_{\tilde{q}}|$.\label{fig:boost}}
\end{minipage}
\end{figure}

Production of strongly interacting particles will have the largest
cross sections at the Large Hadron Collider (LHC). Therefore, in
Ref.~\cite{MoortgatPick:2009jy}, we studied possible CP-violating
effects in squark decay chains
\begin{equation}\label{eq:chain}
\tilde{q} \to \tilde{\chi}_2^0 + q \to \tilde{\chi}_1^0 \ell^+
\ell^- + q \; .
\end{equation}
In the last step of the above decay chain we have the genuine
three-body leptonic decay of the neutralino $\tilde{\chi}^0_2$. In
this decay chain one can construct the $\mathrm{T}_N$-odd triple
product~\cite{Choi:2005gt} of momenta of the final state particles
\begin{equation}\label{eq:triple}
\mathcal{T}= \vec{p}_q \cdot (\vec{p}_{\ell^+} \times
\vec{p}_{\ell^-})\; .
\end{equation}
Using this triple product one can construct a CP-odd asymmetry
\begin{equation}\label{eq:asy}
\mathcal{A}_T =
\frac{N_{\mathcal{T}_+}-N_{\mathcal{T}_-}}{N_{\mathcal{T}_+}+N_{\mathcal{T}_-}}\;
,
\end{equation}
where $N_{\mathcal{T}_+}$ ($N_{\mathcal{T}_-}$) are the numbers of
events for which $\mathcal{T}$, Eq.~(\ref{eq:triple}), is positive
(negative), see also~\cite{Bartl:2004jj}. At the parton level, the
asymmetry due to the phase of the bino mass parameter $M_1 = |M_1|
\mathrm{e}^{\mathrm{i} \phi_1}$ can reach $15\%$ in the neutralino
$\tilde{\chi}_2^0$ rest frame, cf.\ Fig.~\ref{fig:asymmetry}.
However, due to the internal proton structure, particles produced at
the LHC get large, undetermined boosts. As a consequence, the
asymmetry is strongly diluted as can be seen in
Fig.~\ref{fig:boost}, with the maximum value of $\sim 2\%$. This
makes the observation of CP-violating effects very
challenging~\cite{Ellis:2008hq}.

We show that the discovery potential can be greatly increased if one
can reconstruct  momenta of all the particles involved including
those escaping detection~\cite{MoortgatPick:2009jy}. We apply this
technique to the squark decay chain, Eq.~(\ref{eq:chain}), and
analyze the possible enhancement of the signal.

\section{CP asymmetry in the laboratory frame}

\begin{figure}[b!]
\includegraphics[width=190pt,angle=0]{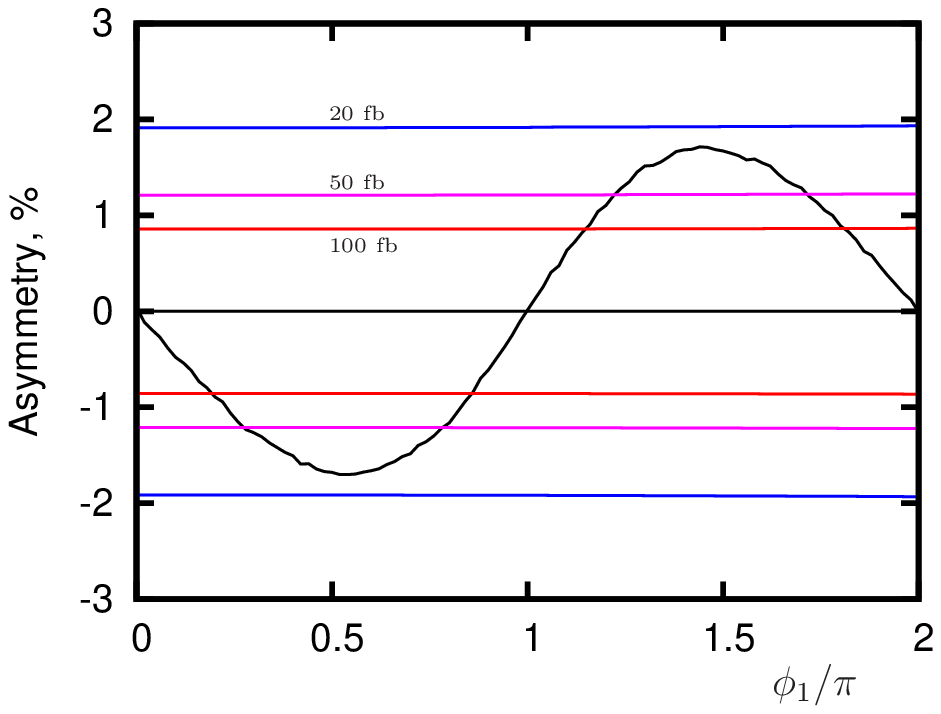}~a)\hfill
\includegraphics[width=190pt,angle=0]{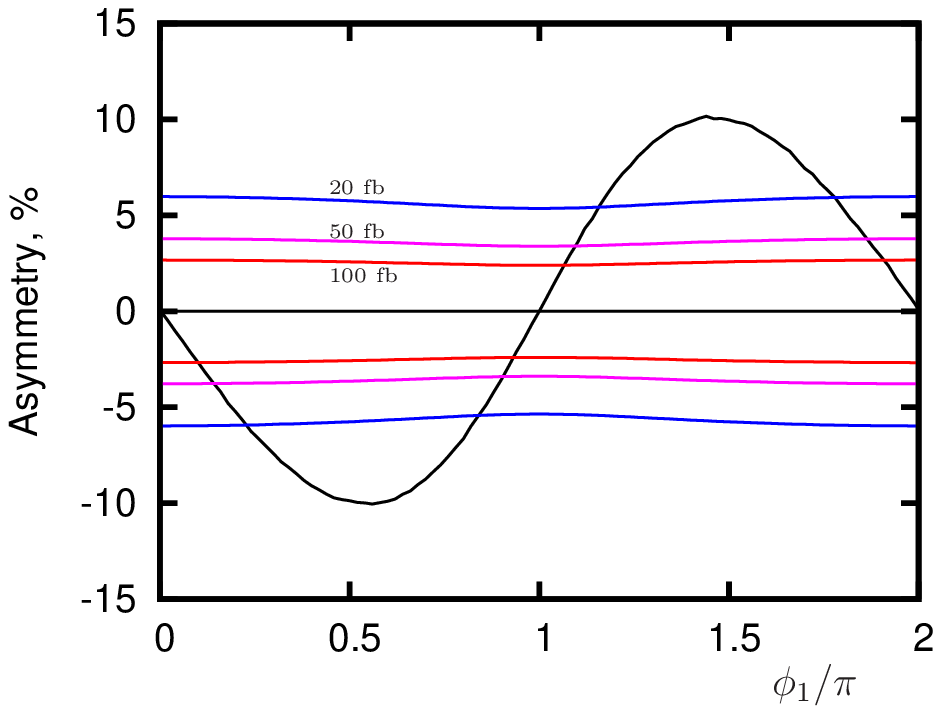}~b) \caption{The
asymmetry $\mathcal{A}_T$ at the LHC with PDFs included in the
analysis: \textbf{(a)} before momentum reconstruction and
\textbf{(b)} after momentum reconstruction. The coloured lines show
the size of the asymmetry needed for a $3\sigma$ observation at the
given luminosity and for $\sqrt{s} = 14$~TeV.\label{fig:results}}
\end{figure}

We start with studying the impact of parton density functions (PDFs)
on the asymmetry. We observe a reduction by an order of magnitude in
the asymmetry, Eq.~\eqref{eq:asy}, compared with the asymmetry in
the neutralino rest frame, see Fig.~\ref{fig:results}a. This is
because in a boosted squark frame the momentum vector of the quark
may now flip to the opposite side of the plane formed by $\ell^+$
and $\ell^-$. This changes the sign of the triple product,
Eq.~(\ref{eq:triple}), causing a significant dilution. The other
dilution factor that has to be taken into account are anti-squarks
$\tilde{q}^*_L$. They will be produced along with squarks, however
at a much lower rate, what is a consequence of the valence quarks
present in the colliding protons~\cite{MoortgatPick:2009jy}. As the
asymmetries due to $\tilde{q}_L$ and $\tilde{q}_L^*$ have the
opposite sign, the asymmetry would vanish if we had equal numbers of
both species.

Including the above effects, we end up with a maximum value of
$|\mathcal{A}_T| = 1.7\%$ for our scenario. Using the total
production cross section and the branching ratios for the decay
chain, Eq.~(\ref{eq:chain}), one can get the expected number of
events. Fig.~\ref{fig:results}a shows the integrated luminosity
needed to observe the asymmetry at the $3\sigma$-level. For
$\mathcal{L} = 100\ \mathrm{fb}^{-1}$ a wide range of values for
$\phi_1$ can be probed.

\section{CP asymmetry with momentum reconstruction}

\begin{figure}
\begin{minipage}{200pt}
\includegraphics[width=190pt,angle=0]{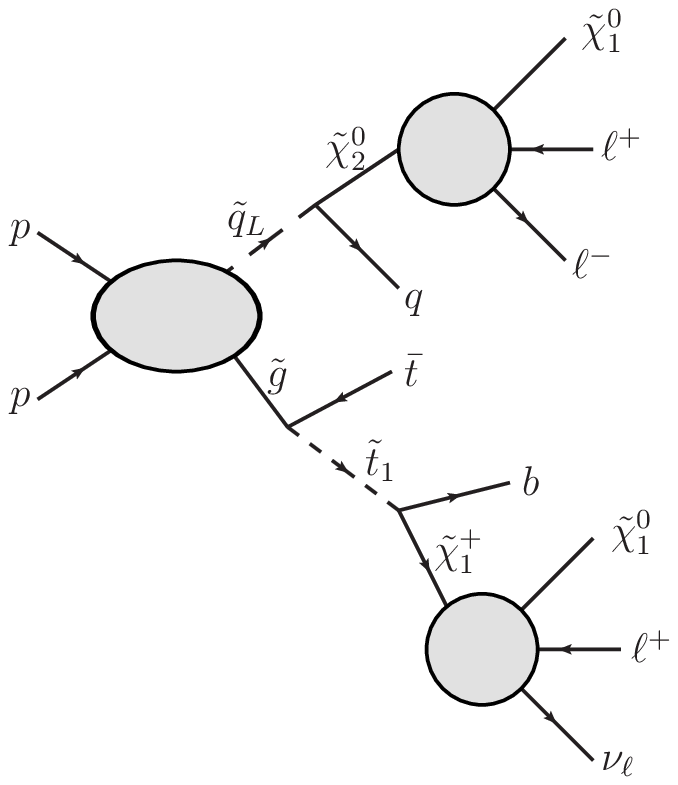}
\caption{The process studied for momentum
reconstruction.\label{fig:fullproc}}
\end{minipage}\hfill
\begin{minipage}{200pt}
\includegraphics[width=190pt,angle=0]{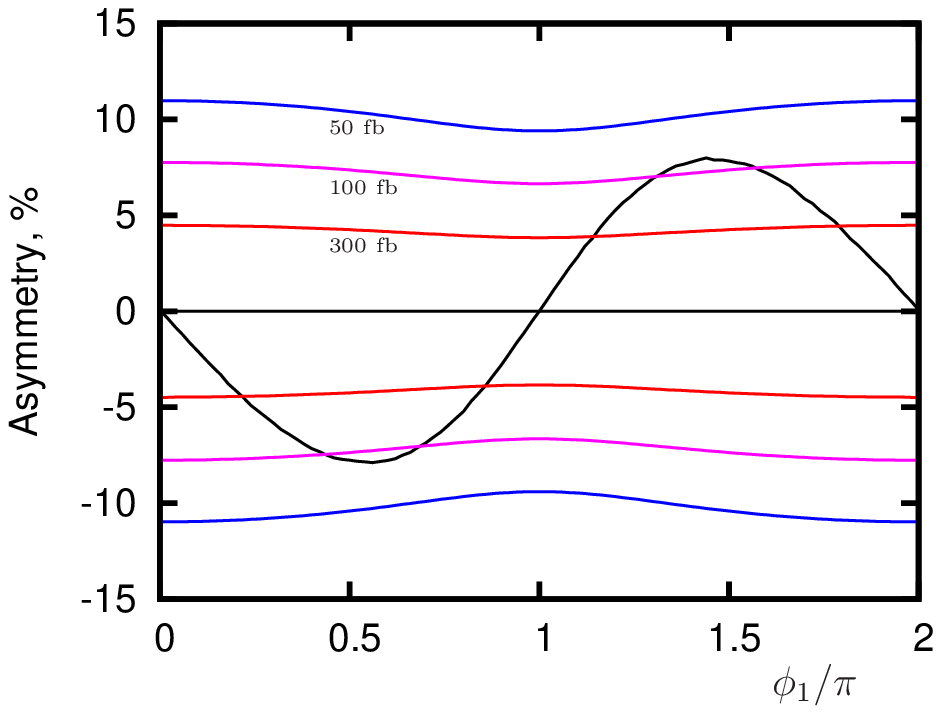}
\caption{The asymmetry $\mathcal{A}_T$ at the LHC after momentum
reconstruction with cuts and momentum smearing. The coloured lines
show the size of the asymmetry needed for a $3\sigma$ observation at
the given luminosity and for $\sqrt{s} = 14$~TeV.\label{fig:momrec}}
\end{minipage}
\end{figure}
As mentioned before the asymmetry has the maximum value in the
neutralino rest frame. Therefore, the full reconstruction of
kinematics of the underlying process would in principle restore the
original value of the asymmetry, cf.~\cite{MoortgatPick:2009jy}.
However, the squark decay chain itself offers too little kinematical
constraints. Hence we consider associated production of squark and
gluino in order to perform the reconstruction,
\begin{equation}
pp \to \tilde{q}_L\; \tilde{g} \;,
\end{equation}
where the following gluino decay chain is dominant
\begin{equation}\label{eq:gluinodecay}
 \tilde{g} \to \tilde{t}_1\; \bar{t} \to \tilde{\chi}^+_1\;  b\;  \bar{t} \to \tilde{\chi}^0_1\; \ell^+\; \nu_{\ell}\;  b\;  \bar{t}\;
 .
\end{equation}
Again, in the last step we have the three-body leptonic chargino
decay, see Fig.~\ref{fig:fullproc}.

For this process one can formulate 6 on-shell~\cite{Kawagoe:2004rz}
conditions for the intermediate particles and the final state LSP
from the squark decay chain:
\begin{IEEEeqnarray}{rCl+rCl}
  m^2_{\tilde{\chi}^0_1} & = & (P_{\tilde{\chi}^0_{1a}})^2 \;,
  &  m^2_{\tilde{\chi}^+_1} & = & (P_{\tilde{\chi}^0_{1b}}+P_{\nu_{\ell}}+P_{\ell^+_b})^2 \;,\IEEEnonumber\\
  m^2_{\tilde{\chi}^0_2} & = & (P_{\tilde{\chi}^0_{1a}}+P_{\ell^+_a}+P_{\ell^-_a})^2 \;,
  & m^2_{\tilde{t}}  & = & (P_{\tilde{\chi}^+_1}+P_b)^2  \;,\\
  m^2_{\tilde{q}} & = & (P_{\tilde{\chi}^0_2}+P_{q})^2 \;,
  & m^2_{\tilde{g}} & = & (P_{\tilde{t}}+P_{t})^2\;. \IEEEnonumber
\end{IEEEeqnarray}
Together with two equations involving missing transverse momentum
\begin{equation}
\overrightarrow{p}^T_{miss} =
\overrightarrow{p}^T_{\tilde{\chi}^0_{1a}} +
\overrightarrow{p}^T_{\tilde{\chi}^0_{1b}} +
\overrightarrow{p}^T_{\nu_{\ell}}\;,  \label{eq:MET}
\end{equation}
this gives 8 equations (6 linear and 2 quadratic). The components of
the four-momenta of the invisible final state particles are the 8
unknowns and the system can therefore be
solved~\cite{MoortgatPick:2009jy}. In principle one gets up to four
real solutions, but there are no additional kinematic constraints to
pick the correct solution. Hence, only the events that give the same
sign for the triple product, Eq.~(\ref{eq:triple}), in all cases,
are taken into account. This guarantees that we take the correct
sign for the triple product for the calculation of the asymmetry but
reduces the number of available events.

The procedure allows one to reconstruct the momenta of the
intermediate particles, in particular $\tilde{q}_L$ and
$\tilde{\chi}^0_2$. It is now possible to calculate the triple
product in the rest frame of the decaying particle, neutralino
$\tilde{\chi}_2^0$, and in principle to restore its maximal
asymmetry, as shown in Fig.~\ref{fig:results}b. There is a
significant improvement compared to the situation before the
reconstruction, cf.\ Fig.~\ref{fig:results}a, but we still observe
dilution due to the anti-squark admixture in the sample.

Finally, we include some of the experimental effects in our
analysis. We impose basic selection cuts and momentum smearing due
to the finite detector resolution, see
Ref.~\cite{MoortgatPick:2009jy} for details. With momentum smearing
included, there is a significant number of events where do not
obtain the correct rest frame of the neutralino $\tilde{\chi}^0_2$.
This results in an increased dilution and a further reduction in the
number of usable events. Nevertheless, the asymmetry can be restored
to a reasonable value in the observable range, see
Fig.~\ref{fig:momrec}.

\section{Conclusions}

In \cite{MoortgatPick:2009jy} we have studied the possibility of
observing CP-violating effects in the squark cascade decay chain at
the LHC. It was shown that CP-odd observables may be successfully
probed by kinematic reconstruction of momenta of the particles
involved in the process. Such an measurement at the LHC can directly
influence future searches at a linear collider and provide hints of
the origin of the matter-antimatter asymmetry of the universe.

\begin{acknowledgement}
KR is supported by the EU Network MRTN-CT-2006-035505 (HEPTools). JT
is supported by the UK Science and Technology Facilities Council
(STFC).
\end{acknowledgement}

\end{document}